\documentclass{PoS}
\usepackage{epsfig}
\title{Soft QCD Results from CDF}

\ShortTitle{Soft QCD Results from CDF}

\author{\speaker{Christina Mesropian}%
         \thanks{for the CDF Collaboration}\\
        Rockefeller University\\
        E-mail: \email{christina.mesropian@cern.ch}}

\abstract{We review
the latest underlying event and minimum bias results and discuss the associated problematics with the idea of establishing a solid baseline
for the LHC experiments. These measurements include study of the underlying event in Drell-Yan and dijet events, 
 and inclusive differential (in $p_T$) cross sections of
centrally ($\mid\eta\mid <$1) produced lambdas, cascades and omegas. We also present recent results on diffraction obtained by the CDF
collaboration. Single-diffractive W and Z production is  discussed. The first experimental observation of exclusive dijets,
exclusive $\chi_{c0}$ mesons, and a search for exclusive diphotons are discussed. We also present results from a study of central rapidity
gap production in soft and hard diffractive events.}

\FullConference{35th International Conference of High Energy Physics\\
		 July 22-28, 2010\\
		 Paris, France}

\begin{document}

\section{Underlying Event and MinBias Studies}
The existence of Monte Carlo models that simulate accurately QCD hard-scattering events is essential for all {\em new} physics searches
 at hadron-hadron colliders. To achieve a given accuracy one should be able not only to have a good model 
of the hard scattering part of the process, but also of the beam-beam remnants (BBR) and the multiple parton interactions (MPI), 
an unavoidable background to most collider observables. 
For Drell-Yan production, the final state includes a lepton-antilepton pair, and there is no colored final state radiation, thus providing a clean way to study the underlying event (UE).
The methodology of the presented study is similar to previous CDF UE studies~\cite{mesropian3-2}, by considering {\em toward}, {\em away}, and {\em transverse} regions defined by the azimuthal angle $\Delta\phi$  relative to  the direction of the leading jet in the event, or  the direction 
of the lepton-pair in Drell-Yan production ($\Delta\phi=\phi-\phi_{jet_1/pair}$). We study charged particles with $p_T>$0.5 GeV/c and $\mid\eta\mid <$1 in the above-mentioned regions. 
For high-$p_T$ jet production we require that the leading jet in the event, reconstructed with the MidPoint algorithm, have $\mid\eta_{jet}\mid<$2. 
For Drell-Yan production we require the invariant mass of the lepton-pair to be in the mass region of the Z-boson, 70$<M_{pair}<$110 GeV/c$^2$, with $\mid\eta_{pair}\mid<$6.
The underlying event observables are found to be reasonably flat with the increasing lepton pair transverse momentum in the {\em transverse} and {\em toward} regions, but 
distributions go up in the {\em away} region to balance lepton pairs. 
\begin{figure}[htbp]
\centerline{\includegraphics[width=0.9\columnwidth]{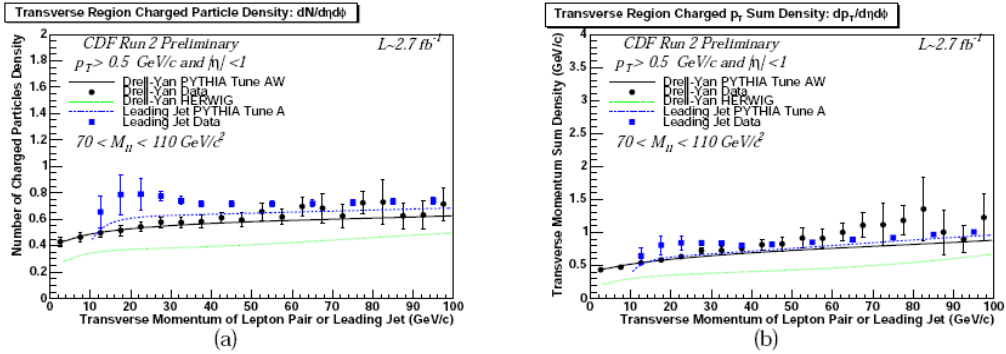}}
  \caption{The underlying event observables in Drell-Yan production.}
\label{mesropian_christina_fig1}
\end{figure}
In Fig.~\ref{mesropian_christina_fig1}(a) and (b), we plot two observables corresponding to the
 underlying event: the charged particle number density and the charged transverse momentum sum density in the transverse region, compared with PYTHIA Tune A 
  and AW~\cite{mesropian3-3}-\cite{mesropian3-4}, HERWIG~\cite{mesropian3-5} without MPI and a previous results of CDF analysis 
of underlying events with the leading jet. There is very good agreement with PYTHIA tune AW MC predictions, while HERWIG produces much less activity. 
The comparison with leading jet underlying event results shows close agreement, which indicates the universality of underlying event modeling.
The behavior of the average charged-particle $p_T$ versus
 charged-particle multiplicity is also important. The rate of
change of $p_T$ versus charged multiplicity is a measure of
the amount of hard versus soft processes contributing, and
it is sensitive to the modeling of the multiple-parton interactions.
PYTHIA Tune A and Tune AW do a good job in
describing the data on $<p_T>$ versus multiplicity for minbias
and Drell-Yan events. The  behavior of $<p_T>$ versus multiplicity is remarkably similar for 
minbias events and Drell-Yan events, suggesting that MPI are playing an important role in both these processes.
Models with MPI predict that the underlying event will become much more active at the LHC. A lot can be learned by comparing the Tevatron results with early measurements at the LHC, to improve modeling of the underlying event for future precision measurements.

\subsection{Hyperons}

We also report a set of measurements of inclusive invariant $p_T$ differential cross sections of $\Lambda^0$, $\bar{\Lambda}^0$, $\Xi^{\pm}$, and $\Omega^{\pm}$ hyperons reconstructed in the central region with $\mid\eta\mid<$1 and $p_T <$10$ GeV/c$. Events were collected with a minimum-bias trigger.
As $p_T$ increases, the slopes of the differential cross sections remain similar, which could indicate universality of particle production in $p_T$.
The production ratios $\Xi^-/\Lambda^0$ and $\Omega^-/\Lambda^0$ were also studied as a function of $p_T$ and are fairly constant in the high $p_T$ region.

\section{Diffractive results}
The CDF collaboration contributed extensively~\cite{mesropian1-1}-\cite{mesropian1-9} to significant progress in understanding diffraction by studying a wide variety of diffractive processes at three different center-of-mass energies: 630 GeV, 1800 GeV, Run I of Tevatron, and 1960 GeV - Run II. Some important results include the observation of  QCD factorization breakdown in hard single diffractive processes, discovery of large rapidity gaps between two jets, study of diffractive structure function  in double pomeron exchange dijet events.
Diffractive W/Z production is an important process for probing the quark content of the pomeron, 
since to leading order, the W/Z is produced through a quark, while  gluon associated production is 
suppressed by a factor of $\alpha_S$ and can be identified by an additional  jet.  
CDF studied diffractive W production in Run I~\cite{mesropian1-2} by using the rapidity gap signature of  diffractive events. 
In Run II, we select events with ``intact leading antiproton'' signature, where $\bar{p}$ is detected in the Roman Pot Spectrometers (RPS).
The RPS allows very precise measurement of the fractional momentum loss of $\bar{p}$  ($\xi$),  eliminating the problem of {\em gap survival probability}.  
The  novel
feature of the analysis, the determination of the full kinematics of the $W\rightarrow l\nu$ decay, is 
made possible by obtaining the neutrino ${E_T}^{\nu}$ from the missing $E_T$, ${E\!\!\!\!/}_T$,  and $\eta_{\nu}$
from the formula ${\xi}^{RPS}-{\xi}^{cal}=\frac{{E\!\!\!\!/}_T}{\sqrt{s}}e^{-{\eta}_{\nu}}$,  
where $\xi^{RPS}$ is the true $\xi$ measured in RPS and ${\xi}^{cal}=\sum_{i(towers)}{(E_T^i/\sqrt{s})exp(-\eta^i)}$.
The fractions of diffractive W and Z events are measured to 
be 
$[0.97\pm0.05(stat.)\pm0.11(syst.)]$\% and $[0.85\pm0.20(stat.)\pm 0.11(syst.)]$\% 
for the kinematic range 0.03$<\xi<$0.10 and $\mid t\mid <$1 GeV/c.
  The measured diffractive W fraction is consistent with the Run I CDF 
result when corrected for the $\xi$ and $t$ range.

\subsection{Central Exclusive production}

The exclusive dijet production was first studied by CDF in Run I data and a limit of 
$\sigma_{excl}<$3.7 nb (95\% CL) was placed~\cite{mesropian2-1}.
This study was continued in Run II when the observation of the exclusive dijet production was reported~\cite{mesropian2-4}.
The exclusive signal is extracted using the dijet mass
fraction  method: 
the
ratio $R_{jj}\equiv M_{jj}/M_X$ of the dijet mass $M_{jj}$ to the total mass
$M_X$ of the final state is formed and
used to discriminate between the signal of exclusive dijets,
defined as $R_{jj}>$0.8, and the background of
inclusive double pomeron exchange dijets, expected to have a continuous distribution
concentrated at lower $R_{jj}$ values. 
The measured cross sections~\cite{mesropian2-4} are consistent with predictions by 
 Khoze et al. ~\cite{mesropian2-5}.

Another process which is closely related to exclusive Higgs production is exclusive
diphoton production $p\bar{p}\rightarrow p\gamma\gamma\bar{p}$. CDF has performed a search for exclusive $\gamma\gamma$   using a 0.5 fb$^{-1}$ data sample obtained with a trigger requiring the presence of two electromagnetic (EM) towers and forward gaps in both forward directions.
Three  candidate
events were found,
by requiring all calorimeters to be empty,
and no tracks to be associated with two EM trigger towers.
Two of these events are likely to be $\gamma\gamma$, and the third is more likely to be $\pi^0\pi^0$. 
A limit was placed on exclusive diphoton production of 410 fb at 95\% CL~\cite{mesropian2-7}. The prediction~\cite{mesropian2-8} is compatible with this limit. CDF plans to update this measurement with additional available data.

CDF II also studied dimuon production, when the event signature requires two oppositely 
charged central muons, and either no other particles (large forward rapidity gaps), or one additional photon detected. 
 Within the kinematic region $\mid\eta(\mu)\mid<$0.6 and $M_{\mu\mu}\in[3.0,4.0]$ GeV/$c^2$, there are 402 events with no 
EM shower, see the $M_{\mu\mu}$ spectrum in Fig.~\ref{mesropian_christina_fig2}(left). 

The $J/\psi$ and $\psi(2S)$ are prominent, together with a continuum. 
By requiring one EM shower with $E_T^{EM}>$80 MeV in addition to the requirement 
mentioned above, we are able to measure $\chi_{c0}$ production. Allowing EM tower 
causes a large increase (+ 66 events) in the $J\/psi$ peak and minor change (+1 event) in 
the $\psi(2S)$ peak. 
 After correcting for background, efficiencies, and the branching fraction, 
we obtain a cross section for exclusive $\chi_{c0}$ production of 75$\pm$10(stat)$\pm$10(syst) nb~\cite{mesropian2-9}, 
which is compatible with the theoretical predictions ~\cite{mesropian2-10}-\cite{mesropian2-12}.

\subsection{Double Diffractive processes}

Double diffractive (DD) dissociation is the process in which 
two colliding hadrons dissociate
into clusters of particles 
(jets in case of hard DD dissociation) producing events with a large 
non-exponentially 
suppressed central pseudo-rapidity gap.
\begin{figure}[htbp]
\centerline{\includegraphics[width=0.40\columnwidth]{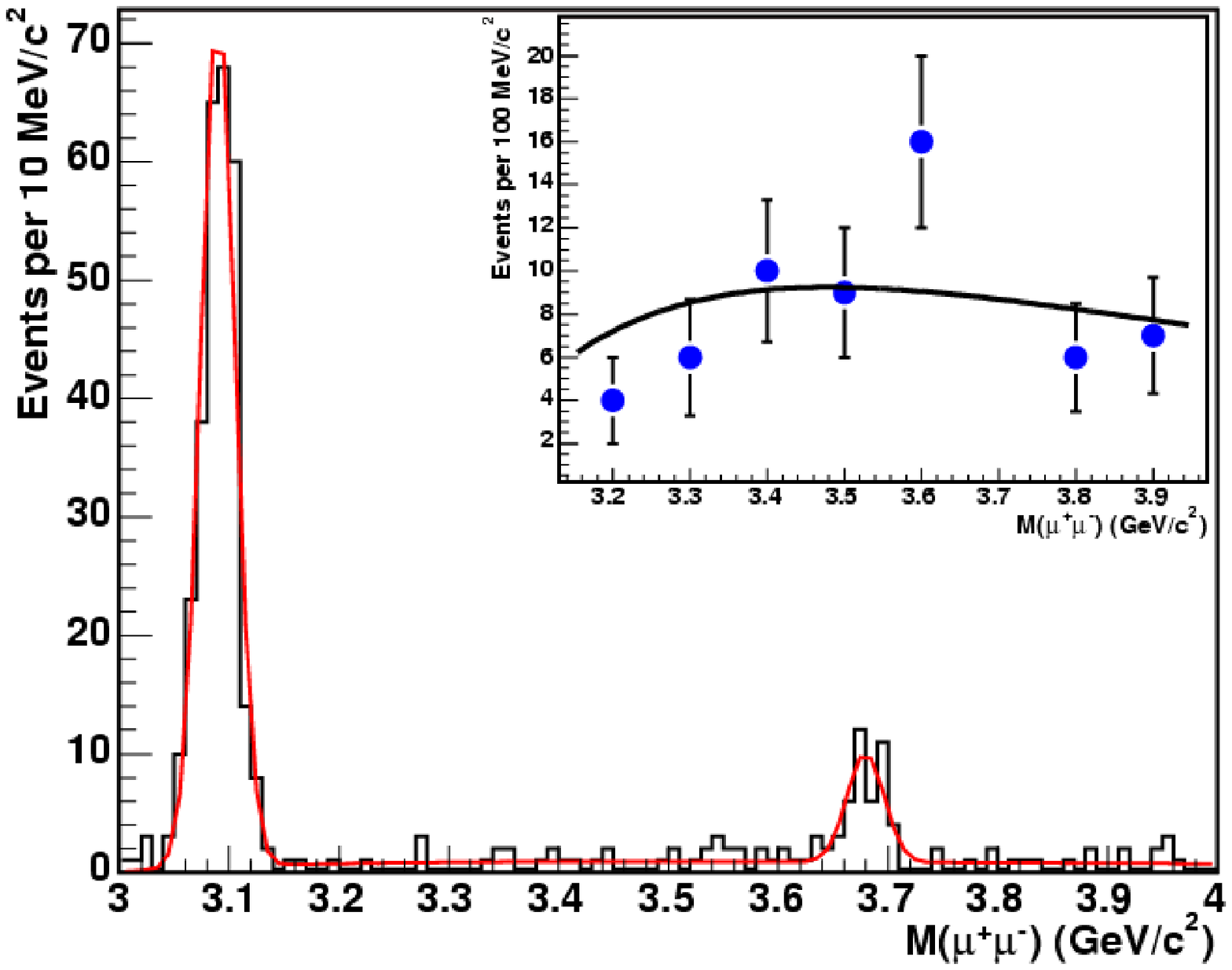}
\includegraphics[width=0.50\columnwidth]{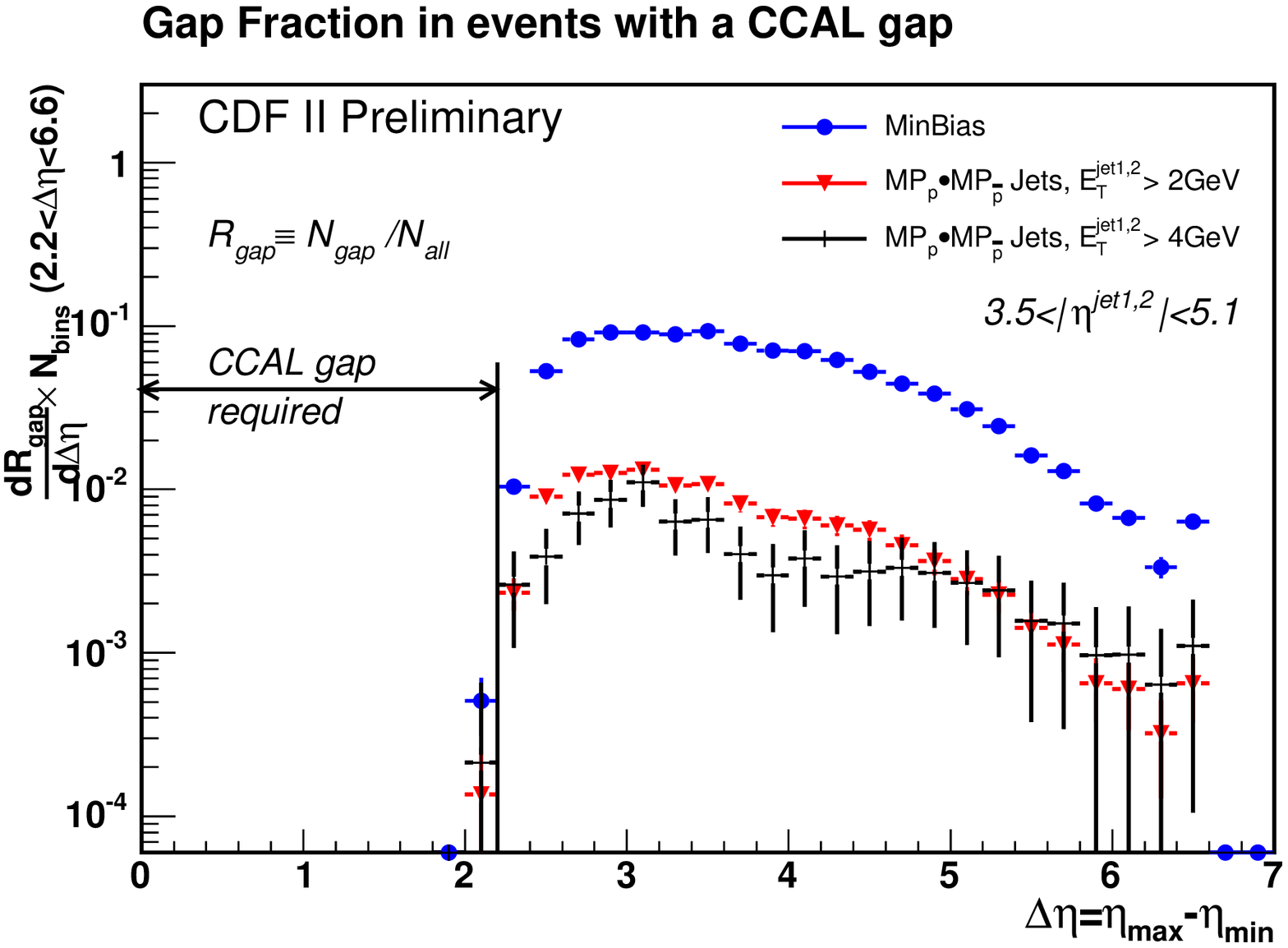}}
  \caption{(left) Mass $M_{\mu\mu}$ distribution of 402 exclusive events, with
no EM shower, (histogram) together with a fit to two Gaussians
for the $J/\psi$  and  $\psi(2S)$, and a QED continuum. All
three shapes are predetermined, with only the normalizations
floating. Inset: Data above the $J/\psi$  and excluding
3.65 $<M_{\mu\mu}<$ 3.75 GeV/c$^2$ ( $\psi(2S)$) with the fit to the QED
spectrum times acceptance (statistical uncertainties only); (right) The distribution of the gap fraction $R_{gap}=N_{gap}/N_{all}$ vs. $\Delta\eta=\eta_{max}-\eta_{min}$ for min-bias  and MiniPlug jet events of $E_T^{jet1,2}>$2 GeV and $E_T^{jet1,2}>$4 GeV.}
\label{mesropian_christina_fig2}
\end{figure}

%
  The extended rapidity coverage provided by the  MiniPlug calorimeters (3.5$<\mid\eta\mid<$5.1) makes CDF II a powerful 
detector for DD studies.
The data sample where jets are required in MP calorimeters  is used to study  ``hard'' diffraction
production.
 ``Soft'' diffractive production is analyzed by examining low 
luminosity data collected with a
min-bias trigger. Fig.~\ref{mesropian_christina_fig2}(right) shows a comparison of the gap fraction,
 as a function of $\Delta\eta$,  between  ``hard'' and ``soft''
 DD production when CCAL gap,
 a rapidity gap  within  -1.1$<\eta<$1.1, is required.
  This comparison is relatively free of systematic uncertainties, 
as detector and beam related effects cancel out. The distributions are similar in shape,  
demonstrating that the gap fraction decreases with increasing $\Delta\eta$ for both ``hard'' and ``soft'' DD productions.

\end{document}